\documentstyle[preprint,aps,eqsecnum,psfig]{revtex}
\begin{document}

\tightenlines
 \preprint{hep-th/0311020}
 \preprint{USTC-ICTS-03-5}
\title{Gauss-Bonnet Black Holes in dS Spaces}

\author{Ron-Gen Cai~\footnote{Email address: cairg@itp.ac.cn}}

\address{Institute of Theoretical Physics, Chinese Academy of Sciences,
   \\
 P.O. Box 2735, Beijing 100080, China\\
 Interdisciplinary Center for Theoretical Study, University of Science \\
  and Technology of China, Hefei, Anhui 230026, China}

\author{Qi Guo~\footnote{Email address: guoqi@itp.ac.cn}}
\address{Institute of Theoretical Physics, Chinese Academy of Sciences,
   \\
 P.O. Box 2735, Beijing 100080, China}

\maketitle
\begin{abstract}
We study the thermodynamic properties associated with black hole
horizon and cosmological horizon for the Gauss-Bonnet solution in
de Sitter space. When the Gauss-Bonnet coefficient is positive, a
locally stable small black hole appears in the case of spacetime
dimension $d=5$, the stable small black hole disappears and the
Gauss-Bonnet black hole is always unstable quantum mechanically
when $d \ge 6$. On the other hand, the cosmological horizon is
found always locally stable independent of the spacetime
dimension. But the solution is not globally preferred, instead the
pure de Sitter space is globally preferred.  When the Gauss-Bonnet
coefficient is negative, there is a constraint on the value of the
coefficient, beyond which the gravity theory is not well defined.
As a result, there is not only an upper bound on the size of black
hole horizon radius at which the black hole horizon and
cosmological horizon coincide with each other, but also a lower
bound depending on the Gauss-Bonnet coefficient and spacetime
dimension. Within the physical phase space, the black hole horizon
is always thermodynamically unstable and the cosmological horizon
is always stable, further, as the case of the positive
coefficient, the pure de Sitter space is still globally preferred.
This result is consistent with the argument that the pure de
Sitter space corresponds to an UV fixed point of dual field
theory.

\end{abstract}

\newpage

\section{Introduction}
Higher derivative curvature terms naturally occur in many
occasions, such as in the quantum field theory in curved
space~\cite{BD} and in the effective low-energy action of string
theories. In the latter case, due to the AdS/CFT
correspondence~\cite{AdS/CFT}, these terms can be viewed as the
corrections of large $N$ expansion of boundary CFTs  in the side
of dual field theory. In the side of gravity, however, because of
the nonlinearity of Einstein equations, it is quite difficult to
find nontrivially exact analytical solutions of the Einstein
equations with these higher derivative terms. In most cases, one
has to adopt some approximation methods or find solutions
numerically.

 Up to the quadratic curvature terms, there is a special
 composition,
 \begin{equation}
 \label{1eq1}
 {\cal L}_{\rm GB}= R_{\mu\nu\gamma\delta}R^{\mu\nu\gamma\delta}
     -4 R_{\mu\nu}R^{\mu\nu} +R^2,
  \end{equation}
  which is often called the Gauss-Bonnet term. The Einstein
  gravity with the Gauss-Bonnet term has some remarkable features
  in some sense. For instance, the resulting equations of motion
 have no more than second derivatives of metric and the theory
 has been shown to be free of ghosts when it is expanded about the
 flat space, evading any problems with unitarity~\cite{BouDes}.
 Further, it has been argued that the Gauss-Bonnet term appears
 as the leading correction \cite{Zwi} to the effective low-energy action of
 the heterotic string theory. In addition, it has already been found that exact analytical
 solutions with spherical symmetry can be obtained in this gravity
 theory~\cite{BouDes,Whee,Zanelli,Cai1}.

 The thermodynamics and geometric structure of the Gauss-Bonnet black hole
 in asymptotically flat space have been analyzed in
 Refs.~\cite{Myers,Wilt}. In a previous paper~\cite{Cai1}~\footnote{The
 thermodynamics and phase structure of black hole solutions
 perturbed by quadratic curvature
 terms in asymptotically AdS space has also been discussed in Refs.~\cite{Norjiri},
 see also \cite{Zanelli} for the case of black holes in the dimensionally
 continued gravity.}, we studied the
thermodynamics and phase structure of topological
 black holes in Einstein gravity with the Gauss-Bonnet term and
 a negative cosmological constant. Those topological black holes
 are asymptotically anti-de Sitter (AdS) and their event horizon
 can be a hypersurface with positive, zero, or negative constant
 curvature. In the present paper, we will study the properties of
 Gauss-bonnet black holes in asymptotically de Sitter (dS) space.
 Studying the Gauss-Bonnet black hole in dS space is of interest in its own
 right. On the other hand, we hope to gain some insights into the dual
 field theory in the sense of the dS/CFT correspondence~\cite{Stro}.

 It is well-known that unlike the cases of asymptotically flat space and
 asymptotically AdS space, it is not an easy matter to calculate
 conserved charges associated with an asymptotically dS space
 because of the absence of spatial infinity and a globally
 timelike Killing vector in such a spacetime. On the other hand,
 there is
a cosmological event horizon, except for the black hole horizon,
for the spacetime of black holes in dS space. Like the black hole
horizon, there is also a thermodynamic feature for the
cosmological horizon~\cite{Haw}. In general the Hawking
temperatures associated with the black hole horizon and
cosmological horizon, respectively, are not equal; therefore the
spacetime for black hole in dS space is unstable quantum
mechanically.

In this paper we will discuss separately the thermodynamics of
black hole horizon and cosmological horizon. Namely, we view the
black hole horizon and cosmological horizon as two thermodynamic
systems. For the case of black hole horizon, we calculate the
black hole mass in the definition due to Abbott and Deser
(AD)~\cite{Abb}, by considering the deviation of metric from the
pure dS space being defined as the vacuum (lowest energy
state)~\footnote{In Ref.~\cite{Abb} the authors consider the
Einstein gravity with a cosmological constant. When higher
derivative curvature terms are present, similar mass definition of
gravitational field has been discussed in Ref.~\cite{Deser}(see
also discussions for Gauss-Bonnet gravity in Ref.~\cite{Pad}).}.
In terms of this definition, the gravitational mass of
asymptotically dS space is always positive, and coincides with the
ADM mass in asymptotically flat space, when the cosmological
constant goes to zero.  For the case of cosmological horizon, we
will adopt the prescription due to Balasubramanian, de Boer and
Minic (BBM)~\cite{BBM}. In this prescription, except for a
constant, which depends on the cosmological constant and space
dimension and can be regarded as the Casimir energy of the dual
field theory in the spirit of the dS/CFT correspondence, the
gravitational mass is just the AD mass, but with an opposite
sign~\cite{BBM,CMZ,Mann,Nojiri2}. The BBM mass is measured at the
far past (${\cal I}^-$) or far further (${\cal I}^+)$ boundary of
dS space, which is outside the cosmological horizon. With these
definitions, thermodynamic quantities associated with the black
hole horizon and cosmological horizon obey  the first law of
thermodynamics, respectively. In Refs.~\cite{Cai2} we have also
shown they satisfy respectively the Cardy-Verlinde formula this
way. In particular, it was argued \cite{Tel} that for the
Euclidean black hole-de Sitter geometry which is closely related
to the horizon thermodynamics, when deals with the thermodynamics
of one of two horizons, one should view the other as the boundary.
In this way, one has well-defined Hamiltonians associated with the
black hole horizon and cosmological horizon, respectively.
Therefore the point of viewing black hole horizon and cosmological
horizon as two thermodynamic systems should be reasonable.

The organization of the paper is as follows. In the next section
we present the solution of the Gauss-Bonnet black hole in dS
space. In Sec.~ III and IV we discuss the thermodynamics and phase
structure of black hole horizon and cosmological horizon,
respectively. This paper is ended in Sec.~V with some conclusions
and discussions.

\section{Gauss-Bonnet black hole solution in de Sitter space}

We start with the Einstein-Hilbert action with the Gauss-Bonnet
term (\ref{1eq1}) and a positive cosmological constant, $\Lambda
=(d-1)(d-2)/2l^2$, in $d$ dimensions
\begin{equation}
\label{2eq1}
 S=\frac{1}{16\pi G}\int d^dx \sqrt{-g}\left (R
 -\frac{(d-1)(d-2)}{l^2} +\alpha {\cal L}_{\rm GB}\right),
 \end{equation}
 where $G$ is the Newton constant and $\alpha $ is the Gauss-Bonnet
  coefficient with dimension
 ($lengh)^2$. From this action we obtain the equations of motion
 \begin{eqnarray}
\label{2eq2} R_{\mu\nu}-\frac{1}{2}g_{\mu\nu}R &=
&-\frac{(d-1)(d-2)}{2l^2}g_{\mu\nu}
  + \alpha \left (\frac{1}{2}g_{\mu\nu}(R_{\gamma\delta\lambda\sigma}
  R^{\gamma\delta\lambda\sigma}-4 R_{\gamma\delta}R^{\gamma\delta}
  +R^2) \right. \nonumber \\
 &&~~~- \left. 2 RR_{\mu\nu}+4 R_{\mu\gamma}R^{\gamma}_{\ \nu}
  +4 R_{\gamma\delta}R^{\gamma\  \delta}_{\ \mu\ \ \nu}
   -2R_{\mu\gamma\delta\lambda}R_{\nu}^{\ \gamma\delta\lambda} \right).
\end{eqnarray}
For the metric  we adopt the following ansatz of spherical
symmetry
\begin{equation}
\label{2eq3} ds^2 = -e^{2\nu}dt^2 +e^{2\lambda}dr^2
+r^2d\Omega_{d-2}^2,
\end{equation}
where $\nu$ and $\lambda$ are functions of $r$ only, and
$d\Omega_{d-2}^2$ represents the line element of a
($d-2$)-dimensional unit sphere with volume
$\Omega_{d-2}=2\pi^{(d-1)/2}/\Gamma[(d-1)/2]$. To find a solution
with metric (\ref{2eq3}), there is a simple method~\cite{BouDes}:
 substituting the metric ansatz (\ref{2eq3}) into the action
(\ref{2eq1}) yields
 \begin{equation}
 \label{2eq4}
 S=\frac{(d-2)\Omega_{d-2}}{16\pi G}\int dt\ dr e^{\nu+\lambda}\left [ r^{d-1}
 \varphi (1+\tilde \alpha \varphi) -\frac{r^{d-1}}{l^2}\right]',
\end{equation}
where a prime denotes the derivative with respect to $r$,
$\tilde\alpha =\alpha (d-3)(d-4)$ and $\varphi =r^{-2}(1-
e^{-2\lambda})$. From the action one has
\begin{eqnarray}
&& e^{\nu +\lambda}=1, \nonumber \\
&& \varphi (1+\tilde\alpha \varphi)-\frac{1}{l^2} =
   \frac{16\pi G M}{(d-2)\Omega_{d-2} r^{d-1}}.
\end{eqnarray}
Then one obtains the exact solution
\begin{equation}
\label{2eq6}
 e^{2\nu} =e^{-2\lambda}=1
+\frac{r^2}{2\tilde\alpha}\left ( 1 \mp
 \sqrt{1+\frac{64 \pi G\tilde \alpha M}{(d-2)\Omega_{d-2} r^{d-1}}+
   \frac{4\tilde\alpha}{l^2}} \right),
\end{equation}
where $M$ is an integration constant, which is just the AD mass of
the solution. This exact solution was first found by Boulware and
Deser in Ref.~\cite{BouDes}. In \cite{Cai1} we extended this
solution to the case where the unit sphere $d\Omega_{d-2}$ is
replaced by a hypersurface with positive, zero or negative
constant curvature.

Note that the solution (\ref{2eq6}) has a singularity at $r=0$ if
$\tilde \alpha >0$. When $\tilde \alpha <0$  there is an
additional singularity at the place where the square root vanishes
in (\ref{2eq6}). In addition note that there are two branches in
the solution (\ref{2eq6}) with  ``$-$" and ``+" signs,
respectively. When the integration constant, $M$, vanishes, the
solution reduces to
\begin{equation}
\label{2eq7}
 e^{2 \nu}=e^{-2\lambda}=1
+\frac{r^2}{2\tilde\alpha}\left ( 1 \mp
 \sqrt{1+ \frac{4\tilde\alpha}{l^2}} \right).
 \end{equation}
 This is a dS or AdS solution depending on the effective
 curvature radius,
 \begin{equation}
 \label{2eq8}
 \frac{1}{l^2_{\rm eff}}=- \frac{1}{2\tilde \alpha} \left (1 \mp
 \sqrt{1+\frac{4\tilde \alpha}{l^2}}\right).
 \end{equation}
 When $\tilde \alpha >0$, one has $l^2_{\rm eff} >0$ for the
 branch with the ``$-$" sign, while $l^2_{\rm eff} <0$ for the
 ``+" sign.  Therefore, in this case, the solution is
 asymptotically dS for the branch with the ``$-$" sign, and
 asymptotically AdS for the sign ``+", although the cosmological
 constant, $\Lambda$, in the action ({\ref{2eq1}) is positive.
 On the other hand, when $\tilde \alpha <0$, one has $l^2_{\rm
 eff} >0$ for both branches, which means the solution is always
 asymptotically dS. But in that case, one can see from
 (\ref{2eq8}) that the Gauss-Bonnet parameter has to satisfy
 \begin{equation}
 \label{2eq9}
 \tilde \alpha /l^2  \ge -1/4,
 \end{equation}
 for the branch with the ``$-$" sign. Otherwise, the theory is not
 well defined. Here it should be stressed that the constraint
 (\ref{2eq9}) is obtained from the vacuum solution of the theory
 (\ref{2eq1}). To avoid a naked singularity, the more stringent
 constraint will be (\ref{3eq19}), as discussed below.

That the solution (\ref{2eq6}) has two branches implies that the
theory has two different vacua (\ref{2eq7}). In Ref.~\cite{BouDes}
Boulware and Deser have shown that the  branch with the ``+" sign
is unstable; the graviton propagating on the background in this
branch is ghost, while the branch with the ``$-$" sign is stable
and the graviton is free of ghost. The branch with the ``+" sign
is of less physical interest. Therefore we will not discuss this
branch and focus on the branch with the ``$-$" sign in what
follows.

\section{Thermodynamics of the black hole horizon}

On can see from (\ref{2eq7}) that when $M=0$, there is a
cosmological horizon at $r_c= l_{\rm eff}$. When $M$ increases
from zero, like the case of the Schwarzschild-dS solution, a black
hole horizon appears in the solution (\ref{2eq6}) and the
cosmological horizon shrinks. That is, in general there are two
positive real roots for the equation, $e^{2\nu}=0$. The large one
is the cosmological horizon $r_c$ and the small one is the black
hole horizon $r_+$. In this section we first discuss the
thermodynamics associated with the black hole horizon.

In terms of the black hole horizon, the mass of the Gauss-Bonnet
black hole, namely the AD mass of the solution, can be expressed
as
\begin{equation}
\label{3eq1}
 M=\frac{(d-2)\Omega_{d-2}r_+^{d-3}}{16\pi G}\left ( 1 +
 \frac{\tilde \alpha}{r_+^2} -\frac{r_+^2}{l^2}\right).
 \end{equation}
 Obviously, when $\tilde \alpha =0$, this quantity reduces to
 the mass of the Schwarzschild-dS black hole in $d$ dimensions.
 The Hawking temperature associated with the black hole horizon
 can easily be obtained by requirement of the absence of conical
 singularity at the black hole horizon in the Euclidean sector of
 the Gauss-Bonnet black hole solution in dS space. It turns out
 \begin{equation}
 \label{3eq2}
 T = \frac{(d-5) \tilde \alpha + (d-3) r_+^2
 -(d-1)r_+^4/l^2}{4\pi r_+ (r_+^2 +2 \tilde \alpha)}.
 \end{equation}
 Another important thermodynamic quantity is the entropy of black
 hole horizon. In Einstein gravity, entropy of black hole
 satisfies the so-called area formula~\cite{GH}. Namely the
 entropy is equal to one-quarter of the horizon area. When higher
 derivative curvature terms are present, however, this statement
 no longer holds.  Wald has shown that entropy of black hole in
 any gravity theory is always a function of horizon
 geometry~\cite{Wald}.  From Refs.~\cite{Cai1,Myers} we can
 read the entropy of the Gauss-Bonnet black hole in dS space
 \begin{equation}
 \label{3eq3}
 S=\frac{\Omega_{d-2}r_+^{d-2}}{4G}\left( 1+
    \frac{2(d-2)\tilde \alpha}{(d-4) r_+^2}\right),
 \end{equation}
 since the cosmological constant does not explicitly occur in this
 expression. Indeed we can show that three thermodynamic
 quantities (\ref{3eq1}), (\ref{3eq2}) and (\ref{3eq3}) obey the
 first law of thermodynamics, $dM =TdS$.

 The quantity indicating the local stability of black hole is the heat capacity.
 For the Gauss-Bonnet black hole in dS space, it is
 \begin{equation}
 \label{3eq4}
 C \equiv \left(\frac{\partial M}{\partial T} \right)= \left(\frac{\partial
 M}{\partial r_+}\right) \left (\frac{\partial r_+}{\partial T}\right ),
 \end{equation}
 where
 \begin{eqnarray}
 \label{3eq5}
 && \frac{\partial M}{\partial r_+} =
    \frac{(d-2)\Omega_{d-2}}{4G } r_+^{d-5}\left(r_+^2 +2 \tilde
    \alpha \right )T, \nonumber \\
 && \frac{\partial T }{\partial r_+}= \frac{1}{4 \pi l^2 r_+^2 (r_+^2 +2 \tilde
          \alpha )^2 }
     [ -(d-1) r_+^6 -(d-3)l^2 r_+^4 -6 (d-1)\tilde \alpha r_+^4
       \nonumber \\
       && ~~~~~~~~~ +2 (d-3)\tilde \alpha l^2 r_+^2 -3(d-5)\tilde
       \alpha l^2 r_+^2 -2 (d-5)\tilde \alpha ^2 l^2].
 \end{eqnarray}
By definition, $F=M-TS$, the free energy of the black hole is
\begin{eqnarray}
\label{3eq6}
 F= && \frac{\Omega_{d-2}r_+^{d-5}}{16\pi G (d-4) l^2 (r_+^2 +2 \tilde
 \alpha) }[(d-4)r_+^6 +(d-4)l^2 r_+^4 \nonumber \\
    && +6 (d-2)\tilde \alpha r_+^4 +(d-8) \tilde \alpha l^2 r_+^2
    + 2(d-2)\tilde \alpha ^2 l^2].
  \end{eqnarray}

Now we are in a position to discuss the thermodynamic stability
and phase structure of the black hole.

(1)  Let us first consider the case of $\tilde \alpha >0$, which
is the case of the heterotic string theory~\cite{Zwi}. When the
Hawking temperature (\ref{3eq2}) vanishes, we obtain
\begin{equation}
\label{3eq7}
 r_+^2=r_{1,2}^2=\frac{(d-3)l^2}{2(d-1)}\left (1\pm
 \sqrt{1+\frac{(d-1)(d-5)}{(d-3)^2}\frac{4\tilde \alpha}{l^2}}
   \right),
\end{equation}
from which we see that when $d=5$ only, there are two real
positive  roots: one is $r_+=r_2=0$, the other is
$r_+^2=r_1^2=l^2/2$. The large one $r_1$ is the horizon radius of
maximal black hole in the solution (\ref{2eq6}), beyond which the
singularity behind the black hole horizon becomes naked. The
maximal black hole with radius $r_1$ in (\ref{3eq7}) is therefore
the counterpart of the Nariai black hole in the Gauss-Bonnet
gravity, there the black holer horizon and cosmological horizon
coincide with each other and therefore the Hawking temperature is
zero.

 When $d=5$, the inverse temperature starts from infinity at
 $r_+=0$,  reaches a minimal value at some place and then goes
 to infinity again at the maximal black hole horizon radius.
 This behavior can be seen from the heat capacity (\ref{3eq5}).
  When the black hole horizon satisfies
 \begin{equation}
 \label{3eq8}
 0< r_+^2 < r_0^2 = \frac{l^2}{4}\left (1+\frac{12\tilde
 \alpha}{l^2}\right) \left( \sqrt{1+\frac{16\tilde\alpha }{l^2}
 \left (1+\frac{12 \tilde \alpha}{l^2}\right)^{-2}}-1\right),
 \end{equation}
the heat capacity is positive and it becomes negative for $r_0^2
 < r_+^2 <l^2/2$. Here $l/\sqrt{2}$ is the maximal black hole
 horizon radius in the case of five dimensions. This behavior is
 quite different from the case without the Gauss-Bonnet term,
 there the heat capacity of the black hole horizon is always
 negative. Therefore the small Gauss-Bonnet black hole satisfying
 (\ref{3eq8})is thermodynamically stable. Here there is no
 restriction on the Gauss-Bonnet coefficient, except for the positivity
 of the coefficient.

 When $d\ge 6$, the equation $T=0$ has only one real positive root $r_1$ in
 (\ref{3eq7}), the other is negative, without any physical
 meaning. One can see from (\ref{3eq2}) that the inverse temperature always starts
 from zero at $r_+=0$ and
 goes to infinity monotonically at the maximal horizon radius
 $r_1$ in
 (\ref{3eq7}), which implies that the heat capacity is always
 negative in this case. This indicates the instability of the
 Gauss-bonnet black hole. When $d \ge 6$, therefore the thermodynamics properties of the
 Gauss-Bonnet black hole in dS space is qualitatively similar to those of
 Schwarzschild-dS black hole, the black hole in dS space
 without the Gauss-Bonnet term. Thus the thermodynamics of
the black hole horizon becomes remarkably related to the spacetime
dimension. In Fig.~1 we plot the inverse temperature
 versus the radius of black hole horizon.

 Checking the free energy (\ref{3eq6}), however, we find that it
 is always positive whatever the spaectime dimension and the
 Gauss-Bonnet coefficient are. In Fig.~2 the free energy of a
 five-dimensional Gauss-Bonnet black hole is plotted versus the
 horizon radius and the Gauss-Bonnet coefficient. In Fig.~3 we plot the free
 energy versus the horizon radius and spacetime dimension with a
 fixed Gauss-Bonnet coefficient. The positivity of the free energy
 implies that the black hole solution is not globally preferred,
 instead the dS space (\ref{2eq7}) is globally preferred since we
 have taken the dS space as the vacuum state.

(2) When $\tilde \alpha <0$, from the solution (\ref{2eq6}) we
find that the black hole horizon must satisfies
\begin{equation}
\label{3eq9}
 r_+^2 \ge -2\tilde \alpha.
 \end{equation}
 However, from the entropy formula ({\ref{3eq3}) one can see
 when
 \begin{equation}
 \label{3eq10}
 -2 \tilde \alpha \le r_+^2 < -\frac{d-2}{d-4} 2 \tilde \alpha,
 \end{equation}
 the entropy of black hole horizon is negative, which should be
 ruled out in the physical phase space since a negative entropy is
 meaningless. Therefore we obtain a
 constraint on the minimal horizon radius of the Gauss-Bonnet
 black hole
 \begin{equation}
 \label{3eq11}
 r_+^2 \ge -2\tilde \alpha \frac{d-2}{d-4},
 \end{equation}
 when $\tilde \alpha <0$. As the above equation holds, the black hole
 has vanishing entropy.

In this case, when $d=5$, from the temperature (\ref{3eq2}), we
find that the horizon radius falls into the range
\begin{equation}
\label{3eq12}
 -6\tilde \alpha \le r_+^2 \le l^2/2.
 \end{equation}
 As the case of $\tilde \alpha >0$, here $r_+ =l/\sqrt{2}$ is the
 horizon radius of the maximal black hole, there both the black
 hole horizon and cosmological horizon coincide with each other
 and the Hawking temperature vanishes. From (\ref{3eq12}) we see
 that there is a more stringent constraint than the one (\ref{2eq9}):
 \begin{equation}
 \label{3in1}
 \tilde \alpha/l^2 \ge -1/12.
 \end{equation}
 Further, one can see from (\ref{3eq5}) that the heat capacity changes its behavior
at the place
 \begin{equation}
 \label{3eq13}
 r_+^2= \frac{l^2}{4} \left(1 + \frac{12\tilde \alpha}{l^2}\right)
 \left ( -1 \pm \sqrt{1 +\frac{16\tilde \alpha}{l^2}\left(1
  +\frac{12 \tilde\alpha}{l^2}\right)^{-2}}\right).
\end{equation}
In order the equation (\ref{3eq13}) to have real root, one has to
have
\begin{equation}
\label{3eq14}
 \tilde \alpha /l^2 \le -1/12,
 \end{equation}
 which contradicts the condition (\ref{3in1}). This means that the inverse
 temperature always starts monotonically  from a finite value at the minimal radius given
 by (\ref{3eq12}) to infinity at the maximal radius $l/\sqrt{2}$ given
 in (\ref{3eq12}). This indicates
 that when $\tilde \alpha <0$, the five-dimensional Gauss-Bonnet black
  hole in dS space has a negative heat capacity and then it is
  unstable as the case without the Gauss-Bonnet term.
 The inverse temperature of the black hole horizon, plotted in Fig.~4,
 shows this fact.

 When $d \ge 6$, the condition that the Hawking temperature (\ref{3eq2})
 vanishes is
 \begin{equation}
 \label{3eq15}
 r_+^2 =r^2_{3,4}= \frac{(d-3)l^2}{2 (d-1)}\left( 1 \pm \sqrt{1+
   \frac{(d-1)(d-5)}{(d-3)^2}\frac{4\tilde \alpha}{l^2}}\right).
 \end{equation}
 To have two positive real roots, one has
 \begin{equation}
 \label{3eq16}
 \frac{\tilde \alpha}{l^2} > -\frac{(d-3)^2}{4(d-1)(d-5)}.
 \end{equation}
 The large root corresponds to the maximal Gauss-Bonnet black hole
in dS space. But the small one is outside the constraint
(\ref{3eq11}). Therefore, the behavior of the inverse temperature
is similar to the case of $d=5$: it starts from a finite value at
the minimal black hole horizon given by (\ref{3eq11}) and goes to
infinity at the maximal radius given (\ref{3eq15}) monotonically.
As a result, the $d\ge 6$ Gauss-Bonnet black hole in dS space is
also unstable when $\tilde \alpha <0$, like the case without the
Gauss-Bonnet term. In Fig.~5 we plot the inverse temperature of
the black hole in seven dimensions with a fixed Gauss-Bonnet
coefficient. Note that the value of the side of right hand in
(\ref{3eq16}) is smaller than $-1/4$ given in ({\ref{2eq9}).
Therefore it seems that the true constraint on the coefficient
$\tilde \alpha$ is (\ref{2eq9}), rather than (\ref{3eq16}). It
turns out this is not correct. The reason is that since the
horizon radius falls into the range
\begin{equation}
\label{3eq18}
 -2\tilde \alpha \frac{d-2}{d-4} \le r_+^2 \le r_3^2,
\end{equation}
which gives us a more stringent constraint
\begin{equation}
\label{3eq19}
 \frac{\tilde\alpha}{l^2}\ge -\frac{(d^2-d-8)(d-4)}{4(d-1)(d-2)^2}.
 \end{equation}
 We have checked numerically  that within the ranges (\ref{3eq18}) and
 (\ref{3eq19}), the heat capacity is always negative, while the free
 energy of the black hole horizon is
 always positive as the case of $\tilde \alpha >0$~\footnote{In the range
 $- 2 \tilde \alpha < r^2_+ < -2 \tilde \alpha (d-2)/(d-4)$, a stable small
 black hole with positive heat capacity may appear, but it has a negative
 entropy. As a result, it should be ruled out in the physical phase
 space, the true constraint on the horizon radius is given by
 (\ref{3eq18}).}.

\section{Thermodynamics of the cosmological horizon}

For the cosmological horizon denoted by $r_c$, the associated
Hawking temperature $T_c$ is
\begin{equation}
\label{4eq1}
 T_c  = \frac{-(d-5) \tilde \alpha - (d-3) r_c^2
 + (d-1)r_c^4/l^2}{4\pi r_c (r_c^2 +2 \tilde \alpha)}.
 \end{equation}
 and entropy $S_c$
 \begin{equation}
 \label{4eq2}
S_c=\frac{\Omega_{d-2}r_c^{d-2}}{4G}\left( 1+
    \frac{2(d-2)\tilde \alpha}{(d-4) r_c^2}\right).
 \end{equation}
 The thermodynamic energy associated with the cosmological
 horizon can be calculated using the BBM prescription~\cite{BBM} (namely,
 the surface counterterm approach).
 In this prescription, it has been found that the BBM mass for
 black holes in dS spaces in Einstein theory is just the negative
 AD mass, see, for example, Refs.\cite{BBM,CMZ,Mann,Tel}, except
 for a constant, which is not relevant to the present discussion.
 For the Gauss-Bonnet black holes in dS space, the BBM
 prescription is also applicable. As the case of Einstein gravity,
 it turns out that the thermodynamic energy of the cosmological horizon
 is  the negative AD mass (see also \cite{Nojiri2,Norjiri}),
 \begin{equation}
 \label{4eq3}
 E = -M =- \frac{(d-2)\Omega_{d-2}r_c^{d-3}}{16\pi G}\left ( 1 +
 \frac{\tilde \alpha}{r_c^2} -\frac{r_c^2}{l^2}\right).
 \end{equation}
 A self-consistency  check is that these three thermodynamic
 quantities obey the first law of thermodynamics, $dE =T_cdS_c$.
 To see the thermodynamic stability, we calculate the heat
 capacity of the cosmological horizon
\begin{equation}
\label{4eq4}
 C_c \equiv \left (\frac{\partial E}{\partial T_c} \right)
    =\left(\frac{\partial E}{\partial r_c}\right)
    \left(\frac{\partial r_c}{\partial T_c}\right),
\end{equation}
where
\begin{eqnarray}
\label{4eq5}
 && \left(\frac{\partial E}{\partial r_c}\right)=
\frac{(d-2)\Omega_{d-2}}{4G } r_c^{d-5}\left(r_c^2 +2 \tilde
    \alpha \right )T_c, \nonumber \\
 && \frac{\partial T_c }{\partial r_c}= \frac{1}{4 \pi l^2 r_c^2 (r_c^2 +2 \tilde
          \alpha )^2 }
     [ (d-1) r_c^6 +(d-3)l^2 r_c^4 +6 (d-1)\tilde \alpha r_c^4
       \nonumber \\
       && ~~~~~~~~~ -2 (d-3)\tilde \alpha l^2 r_c^2 +3(d-5)\tilde
       \alpha l^2 r_c^2 + 2 (d-5)\tilde \alpha ^2 l^2].
 \end{eqnarray}
 And the free energy, $F_c = E-T_cS_c$, is
 \begin{eqnarray}
 \label{4eq6}
F_c= && \frac{\Omega_{d-2}r_c^{d-5}}{16\pi G (d-4) l^2 (r_c^2 +2
\tilde
 \alpha) }[-(d-4)r_c^6 -(d-4)l^2 r_c^4 \nonumber \\
    && -6 (d-2)\tilde \alpha r_c^4 -(d-8) \tilde \alpha l^2 r_c^2
    - 2(d-2)\tilde \alpha ^2 l^2].
  \end{eqnarray}
  The cosmological horizon radius has a range in size: the minimal value
  is just the maximal black hole horizon $r_3$ in(\ref{3eq15}) discussed
  in the previous section, while the maximal radius is $l_{\rm
  eff}$ given in (\ref{2eq8}), there the integration constant $M$
  vanishes. Namely the cosmological horizon is in the following
  region
  \begin{equation}
  \label{4eq7}
  r^2_3
   \le r_c^2 \le l^2_{\rm eff}.
  \end{equation}
  Within this region, it is easy to show that the heat capacity
  ({\ref{4eq5}) is
  positive and therefore the inverse temperature
  always starts from infinity, where the cosmological horizon
  coincides with the black hole horizon, and monotonically goes to a finite
  value, which corresponds to the inverse temperature of the
  vacuum dS space (\ref{2eq7}). This implies that the
  thermodynamics of the cosmological horizon is locally stable. From
  (\ref{4eq6}) we see that the free energy is always negative.
  But this does not mean that the solution is globally preferred
  since when we calculate the gravitational mass, the pure dS
  space (\ref{2eq7}) is regarded as the vacuum. This vacuum has
  zero AD mass, but has non-zero Hawking temperature and entropy
  associated with the cosmological horizon
  \begin{equation}
  \label{4eq8}
  T_c^{\rm vac}=\frac{1}{2\pi l_{\rm eff}}, \ \ \
  S_c^{\rm vac}=\frac{\Omega_{d-2}l^{d-2}_{\rm eff}}{4G}
   \left (1+\frac{2 (d-2)\tilde \alpha}{(d-4) l^2_{\rm
   eff}}\right).
 \end{equation}
 And the corresponding free energy is
 \begin{equation}
 \label{4eq9}
 F_c^{\rm vac}=-\frac{\Omega_{d-2}l^{d-3}_{\rm eff}}{8\pi G}
\left (1+\frac{2 (d-2)\tilde \alpha}{(d-4) l^2_{\rm
   eff}}\right).
\end{equation}
To see whether or not the solution with non-vanishing $M$ is
globally preferred, we have to compare the two free energies
(\ref{4eq6}) and (\ref{4eq9}):
\begin{equation}
\triangle F =F_c -F_c^{\rm vac}.
\end{equation}
If $\triangle F>0$, the solution with non-vanishing $M$ is not
globally preferred, otherwise it is preferred. It seems difficult
to analytically prove $\triangle F >0$, but we have checked
numerically that indeed $\triangle F
>0$ within the range (\ref{4eq7}). Therefore the pure dS space
(\ref{2eq7}) is globally preferred. Namely although the
thermodynamics of the cosmological horizon is locally stable, it
will decay to the pure dS space.  In Fig.~6 we plot the difference
of the two free energies, $\triangle F$, versus the Gauss-Bonnet
coefficient and the cosmological horizon radius in the case of
five dimensions. The case of ten dimensions is plotted in Fig.~7.

When $\tilde\alpha <0$,  the coefficient has to satisfy the
constraint (\ref{3eq19}), again. Within the horizon range
(\ref{4eq7}) and the constraint (\ref{3eq19}), we have numerically
checked that the difference of the free energies associated with
the cosmological horizon is always positive as the case $\tilde
\alpha >0$ (As an example, we plot the free energy difference
associated with the cosmological horizon in the case of five
dimensions in Fig.~8.).  As a result, the pure de Sitter space
(\ref{2eq7}) is globally preferred again.

\section{Conclusion and Discussion}

In summary we have discussed the thermodynamic properties and
phase structures associated with the black hole horizon and
cosmological horizon for the Gauss-Bonnet black hole-de Sitter
spacetime. The black hole horizon and cosmological horizon are
viewed as two thermodynamic systems. When the Gauss-Bonnet
coefficient is positive, which is the case for the effective low
energy action of the heterotic string theory, a locally stable
small black hole whose radius satisfying (\ref{3eq8}) appears in
$d=5$ dimensions, which is absent in the case without the
Gauss-Bonnet term. When the spacetime dimension $d\ge 6$, the
stable small black hole disappears; the black hole is always
unstable thermodynamically as the case without the Gauss-Bonnet
term. Contrary to the black hole horizon, the cosmological horizon
is always thermodynamically stable with positive heat capacity.
But the Gauss-Bonnet black hole solution in de Sitter space is not
globally preferred, instead the pure de Sitter space (\ref{2eq7})
is globally preferred, which has lower free energy than the case
with nonvanishing $M$.

 One the other hand, when the Gauss-Bonnet coefficient is negative,
there is a bound on the coefficient given by (\ref{2eq9}),
otherwise, the gravity theory is not well-defined (note that the
constraint (\ref{2eq9}) is derived from the vacuum solution of the
theory, it does not warrant that a naked singularity does not
occur in this case. In fact, a true constraint is (\ref{3eq19}),
under which a black hole solution is meaningful.). In this case,
the horizon radius of the Gauss-Bonnet black hole has not only an
upper bound $r_3$ given by (\ref{3eq15}), there the black hole
horizon coincides with the cosmological horizon, but also a lower
bound. From the solution (\ref{2eq6}), the lower bound seems to be
$-2 \tilde \alpha$ given by (\ref{3eq9}). Checking the entropy
(\ref{3eq3}) of black hole horizon tells us that within the range
(\ref{3eq10}), the entropy associated with the black hole horizon
is negative. As a result this range (\ref{3eq10}) should be ruled
out in the physical phase space. Therefore the true lower bound of
the horizon radius is given by (\ref{3eq11}). Further it gives
more stringent constraint on the value of the Gauss-Bonnet
coefficient (\ref{3eq19}). Within the coefficient (\ref{3eq19})
and the horizon range (\ref{3eq18}), the black hole horizon
becomes always thermodynamically unstable and the cosmological
horizon is still thermodynamically stable, that is, in this case
the stable small black hole in five dimensions disappears.
Checking the free energies associated with the black hole horizon
and cosmological horizon, respectively, reveals that the pure de
Sitter space is still globally preferred.

Therefore, both thermodynamic discussions of black hole horizon
and cosmological horizon lead to the same conclusion that a pure
de Sitter space is globally preferred. This result is consistent
with the argument that a pure de Sitter space corresponds to an UV
fixed point of the renormalization group flow of the dual field
theory in the dS/CFT correspondence~\cite{Strom,BBM}.

Finally we would like to stress that as argued in INTRODUCTION,
although a black hole-de Sitter spacetime is unstable quantum
mechanically because two Hawking temperatures associated with
black hole horizon and cosmological horizon are in general not
equal, except for the Nariai solution or its generalizations,
where two temperatures equal to each other. So it is not an easy
matter to study the thermodynamic properties of spacetime for a
black hole in dS space as an entire. In particular, Teitelboim
recently argued \cite{Tel} that for the Euclidean black hole-de
Sitter geometry which is closely related to the horizon
thermodynamics, when deals with the thermodynamics of one of two
horizons, one should view the other as the boundary. In this way,
one has well-defined Hamiltonians associated with the black hole
horizon and cosmological horizon, respectively. In this paper we
just followed this spirit to discuss the thermodynamic properties
associated with black hole horizon and cosmological horizon,
respectively, and to obtain the conclusion that the pure de Sitter
space is globally preferred and it is end point of decay. In
addition, the local stability analysis of black hole horizon and
cosmological horizon might be less motivated just due to different
temperatures. However, when two horizons separate with a very
large distance, the effect of the Hawking evaporation of one
horizon could be negligible on the other horizon. In this sense it
might be of some interest and be of meaning to discuss local
stability of two horizons, respectively. We wish that the present
investigation together with a lot of existing literature
concerning black hole-de Sitter spacetimes is in the way to
completely understand classical and quantum properties of
asymptotically de Sitter spaces.

\section*{Acknowledgments}
This work was initiated while one of authors (R.G.C) was visiting
the ICTS at USTC, whose hospitality is gratefully acknowledged.
The research was supported in part by a grant from Chinese Academy
of Sciences, a grant No. 10325525 from NSFC, and by the Ministry
of Science and Technology of China under grant No. TG1999075401.


\begin{references}
\bibitem{BD}N.~D.~Birrell and P.~C.~Davies, ``Quantum Fields In Curved Space,''
 Cambridge Univ. Press,1982.
\bibitem{AdS/CFT} J.~Maldacena,
Adv.\ Theor.\ Math.\ Phys.\  {\bf 2}, 231 (1998) [Int.\ J.\
Theor.\ Phys.\  {\bf 38}, 1113 (1998)] [hep-th/9711200];
 S.~S.~Gubser, I.~R.~Klebanov and A.~M.~Polyakov,
Phys.\ Lett.\ B {\bf 428}, 105 (1998) [hep-th/9802109];
 E.~Witten,
Adv.\ Theor.\ Math.\ Phys.\  {\bf 2}, 253 (1998) [hep-th/9802150].

\bibitem{BouDes}D.~G.~Boulware and S.~Deser,
Phys.\ Rev.\ Lett.\  {\bf 55}, 2656 (1985).

\bibitem{Zwi}B.~Zwiebach,
Phys.\ Lett.\ B {\bf 156}, 315 (1985);
R.~I.~Nepomechie,
Phys.\ Rev.\ D {\bf 32}, 3201 (1985).


\bibitem{Whee}J.~T.~Wheeler,
Nucl.\ Phys.\ B {\bf 268}, 737 (1986).

\bibitem{Zanelli}J.~Crisostomo, R.~Troncoso and J.~Zanelli,
Phys.\ Rev.\ D {\bf 62}, 084013 (2000) [arXiv:hep-th/0003271];
R.~Aros, R.~Troncoso and J.~Zanelli,
Phys.\ Rev.\ D {\bf 63}, 084015 (2001) [arXiv:hep-th/0011097].

\bibitem{Cai1}R.~G.~Cai,
Phys.\ Rev.\ D {\bf 65}, 084014 (2002) [arXiv:hep-th/0109133].

\bibitem{Myers}R.~C.~Myers and J.~Z.~Simon,
Phys.\ Rev.\ D {\bf 38}, 2434 (1988).

\bibitem{Wilt}D.~L.~Wiltshire,
Phys.\ Rev.\ D {\bf 38}, 2445 (1988).

\bibitem{Norjiri}S.~Nojiri and S.~D.~Odintsov,
Phys.\ Lett.\ B {\bf 521}, 87 (2001) [Erratum-ibid.\ B {\bf 542},
301 (2002)] [arXiv:hep-th/0109122];
S.~Nojiri, S.~D.~Odintsov and S.~Ogushi,
Phys.\ Rev.\ D {\bf 65}, 023521 (2002) [arXiv:hep-th/0108172];
M.~Cvetic, S.~Nojiri and S.~D.~Odintsov,
Nucl.\ Phys.\ B {\bf 628}, 295(2002) [arXiv:hep-th/0112045];
Y.~M.~Cho and I.~P.~Neupane,
Phys.\ Rev.\ D {\bf 66}, 024044 (2002) [arXiv:hep-th/0202140];
I.~P.~Neupane,
Phys.\ Rev.\ D {\bf 67}, 061501 (2003) [arXiv:hep-th/0212092];
I.~P.~Neupane,
arXiv:hep-th/0302132.



\bibitem{Abb}L.~F.~Abbott and S.~Deser,
Nucl.\ Phys.\ B {\bf 195}, 76 (1982).


\bibitem{Stro}A.~Strominger,
JHEP {\bf 0110}, 034 (2001) [arXiv:hep-th/0106113].

\bibitem{Haw}G.~W.~Gibbons and S.~W.~Hawking,
Phys.\ Rev.\ D {\bf 15}, 2738 (1977).

\bibitem{Deser}S.~Deser and B.~Tekin,
Phys.\ Rev.\ Lett.\  {\bf 89}, 101101 (2002)
[arXiv:hep-th/0205318];
S.~Deser and B.~Tekin,
Phys.\ Rev.\ D {\bf 67}, 084009 (2003) [arXiv:hep-th/0212292].

\bibitem{Pad}A.~Padilla,
Class.\ Quant.\ Grav.\  {\bf 20}, 3129 (2003)
[arXiv:gr-qc/0303082];
J.~P.~Gregory and A.~Padilla,
Class.\ Quant.\ Grav.\  {\bf 20}, 4221 (2003)
[arXiv:hep-th/0304250].


\bibitem{BBM}V.~Balasubramanian, J.~de Boer and D.~Minic,
Phys.\ Rev.\ D {\bf 65}, 123508 (2002) [arXiv:hep-th/0110108].

\bibitem{CMZ}R.~G.~Cai, Y.~S.~Myung and Y.~Z.~Zhang,
Phys.\ Rev.\ D {\bf 65}, 084019 (2002) [arXiv:hep-th/0110234].

\bibitem{Mann}A.~M.~Ghezelbash and R.~B.~Mann,
JHEP {\bf 0201}, 005 (2002) [arXiv:hep-th/0111217].


\bibitem{Cai2}R.~G.~Cai,
Phys.\ Lett.\ B {\bf 525}, 331 (2002) [arXiv:hep-th/0111093];
R.~G.~Cai,
Nucl.\ Phys.\ B {\bf 628}, 375 (2002) [arXiv:hep-th/0112253].

\bibitem{Nojiri2}M.~Cvetic, S.~Nojiri and S.~D.~Odintsov,
Nucl.\ Phys.\ B {\bf 628}, 295 (2002) [arXiv:hep-th/0112045].

\bibitem{Tel}C.~Teitelboim,
arXiv:hep-th/0203258;
A.~Gomberoff and C.~Teitelboim,
Phys.\ Rev.\ D {\bf 67}, 104024 (2003) [arXiv:hep-th/0302204].

\bibitem{GH}G.~W.~Gibbons and S.~W.~Hawking,
Phys.\ Rev.\ D {\bf 15}, 2752 (1977).

\bibitem{Wald}R.~M.~Wald,
Phys.\ Rev.\ D {\bf 48}, 3427 (1993) [arXiv:gr-qc/9307038].

\bibitem{Strom}A.~Strominger,
JHEP {\bf 0111}, 049 (2001) [arXiv:hep-th/0110087].




\begin{figure}
\psfig{file=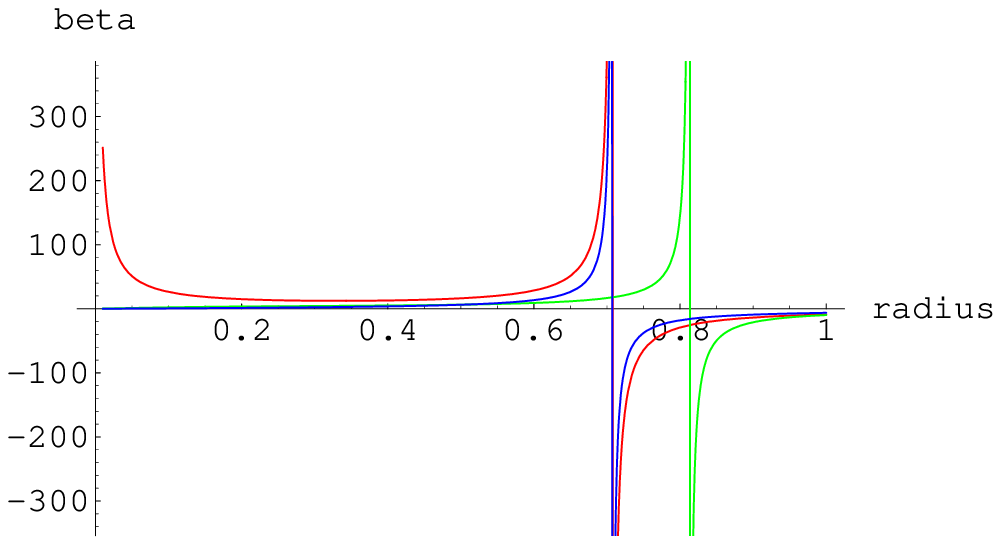,height=80mm,width=100mm,angle=0}
 \caption{The inverse temperature of the Gauss-Bonnet black holes in
 dS space. The red curve corresponds to the case of $d=5$ and
 $\tilde\alpha /l^2=0.2$,  the blue one to the case of $d=5$ and $\tilde\alpha /l^2=0$,
 and the green one to the case of  $d=6$ and
  $\tilde\alpha /l^2=0.2$.  Note that the region with negative temperature
  should be ruled out in the physical phase space.}

\vspace*{2.cm}

 \psfig{file=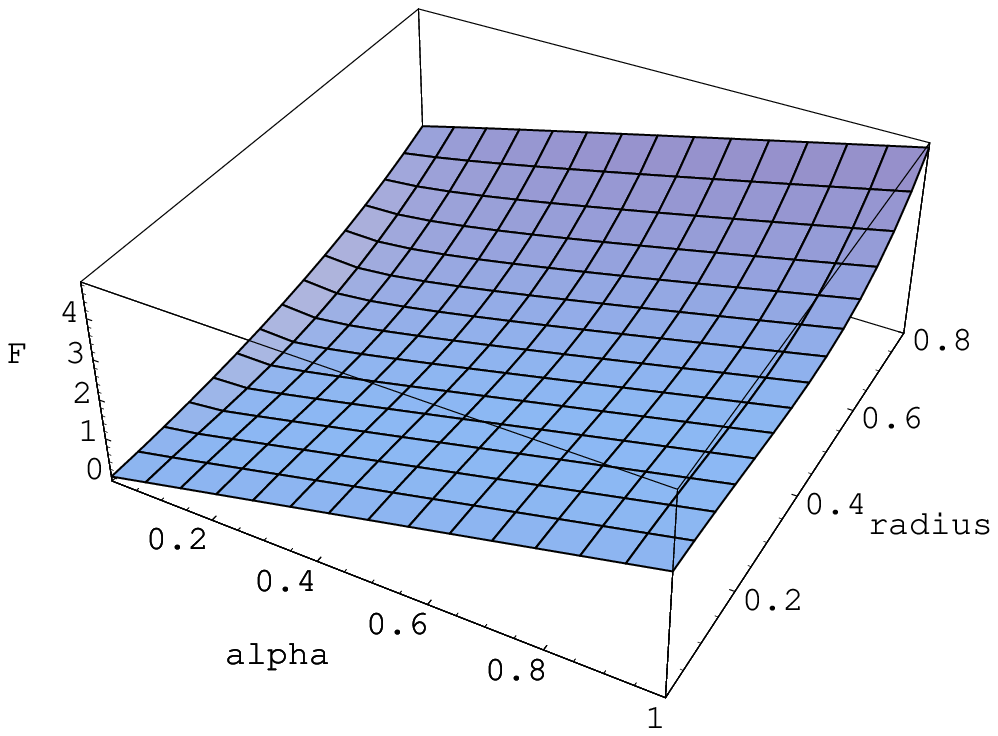,height=80mm,width=100mm,angle=0}
 \caption{The free energy of the five dimensional Gauss-Bonnet black hole versus the
 Gauss-Bonnet coefficient and horizon radius.}

\psfig{file=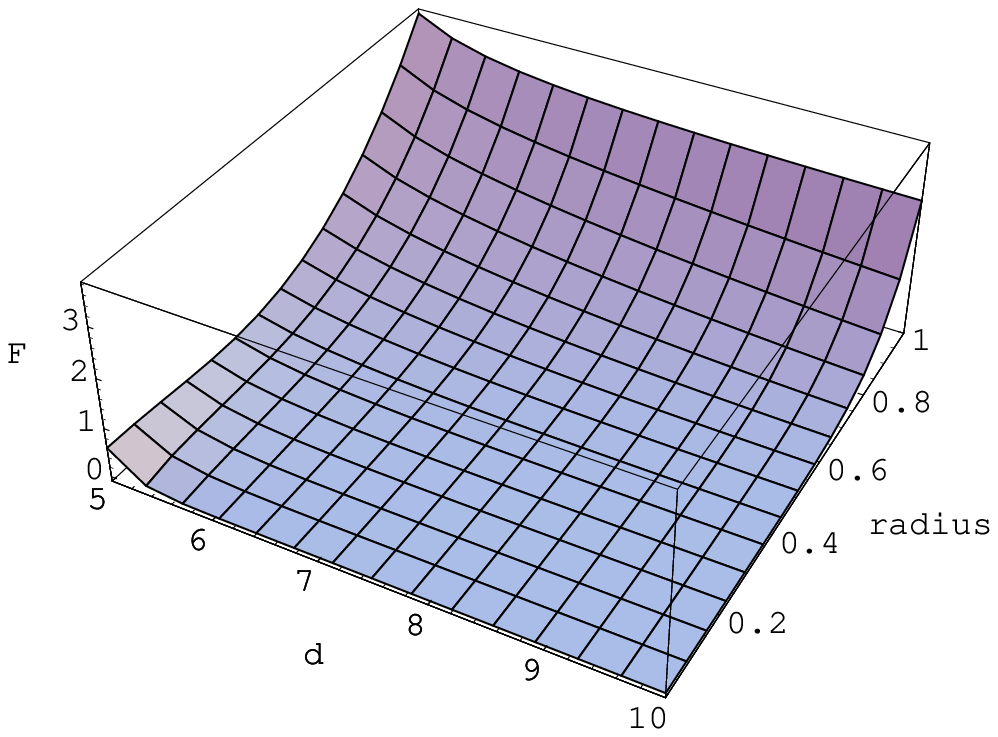,height=80mm,width=100mm,angle=0}
 \caption{The free energy of the Gauss-Bonnet black holes versus
 the horizon radius and spacetime dimension with a fixed
 Gauss-Bonnet coefficient $\tilde \alpha/l^2=0.2$.}

\vspace*{2.cm}

 \psfig{file=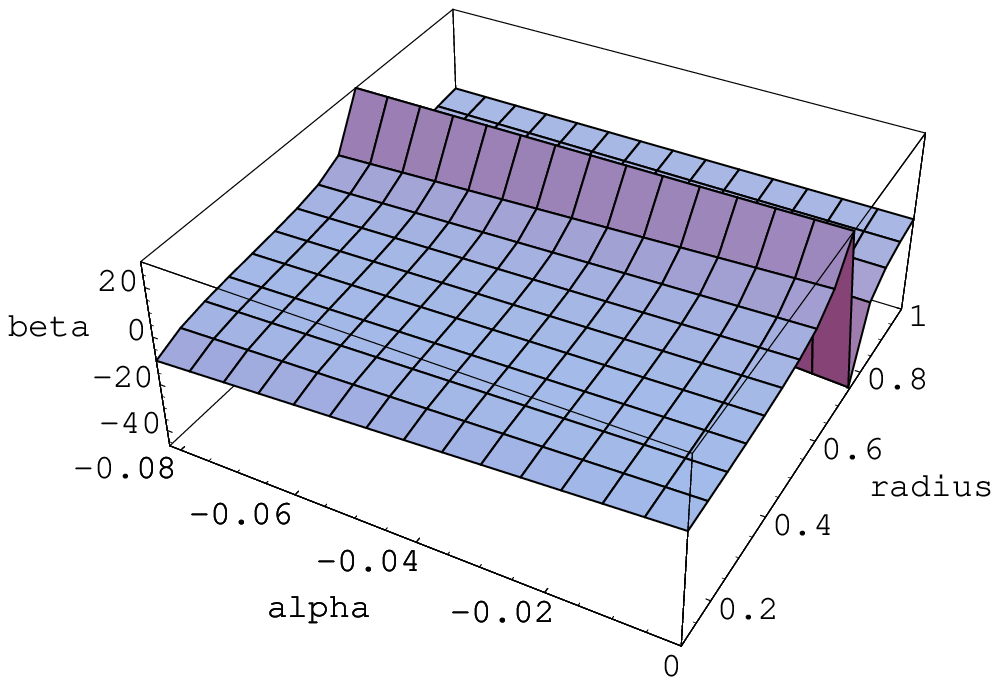,height=80mm,width=100mm,angle=0}
 \caption{The inverse temperature of the five dimensional
  Gauss-Bonnet black holes with $-1/12 <\tilde \alpha/l^2<0$.}

\psfig{file=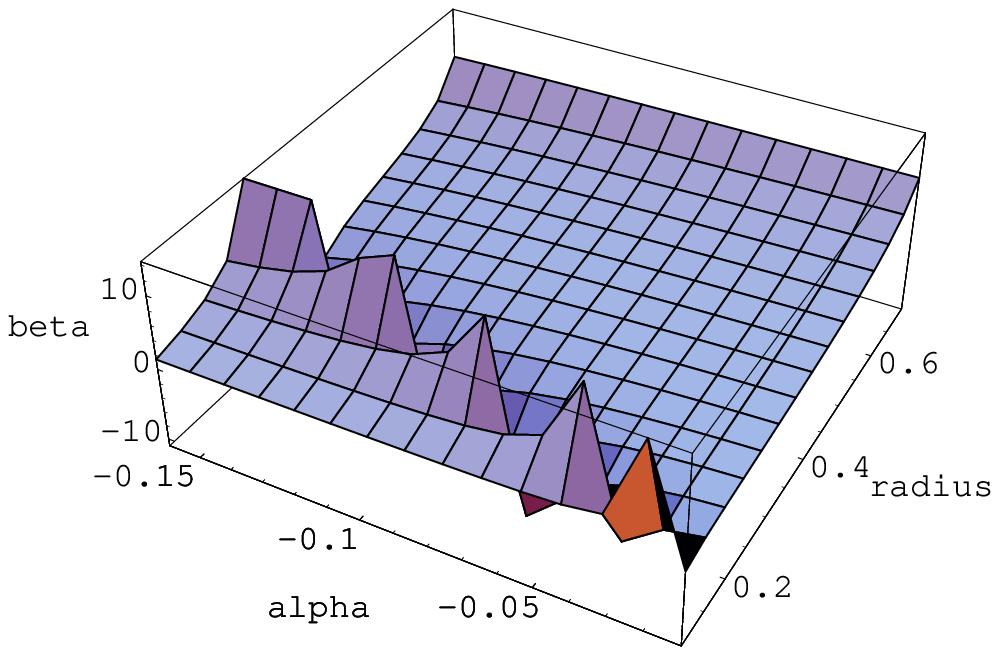,height=80mm,width=100mm,angle=0}
 \caption{The inverse temperature of the seven dimensional
  Gauss-Bonnet black holes with $-17/100 <\tilde \alpha/ l^2 <0$.}

\vspace*{2.cm}

\psfig{file=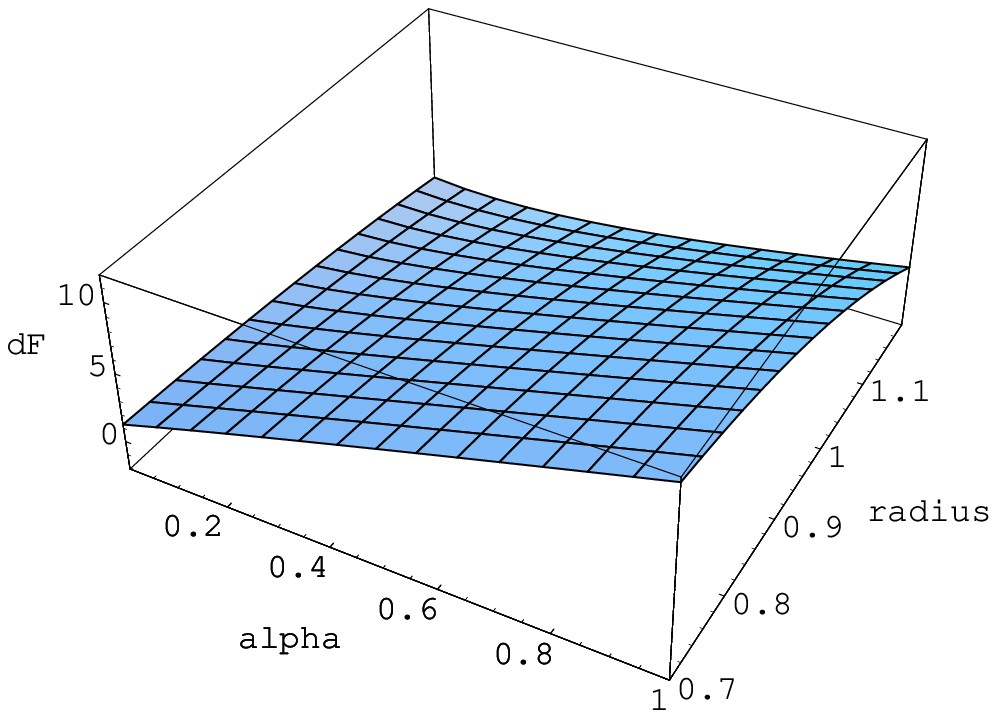,height=80mm,width=100mm,angle=0}
 \caption{The difference $\triangle F$ of two free energies associated with
 the cosmological horizon for the case of five dimensions.}

 \psfig{file=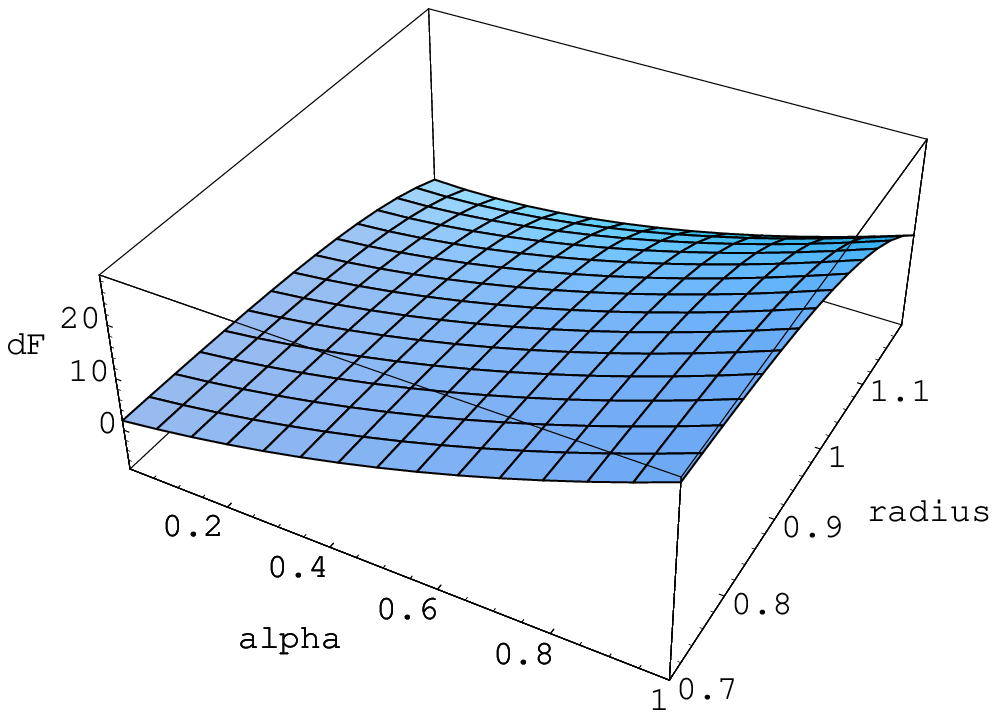,height=80mm,width=100mm,angle=0}
 \caption{The difference $\triangle F$ of two free energies for the
 case of ten dimensions.}


\vspace*{2.cm}

 \psfig{file=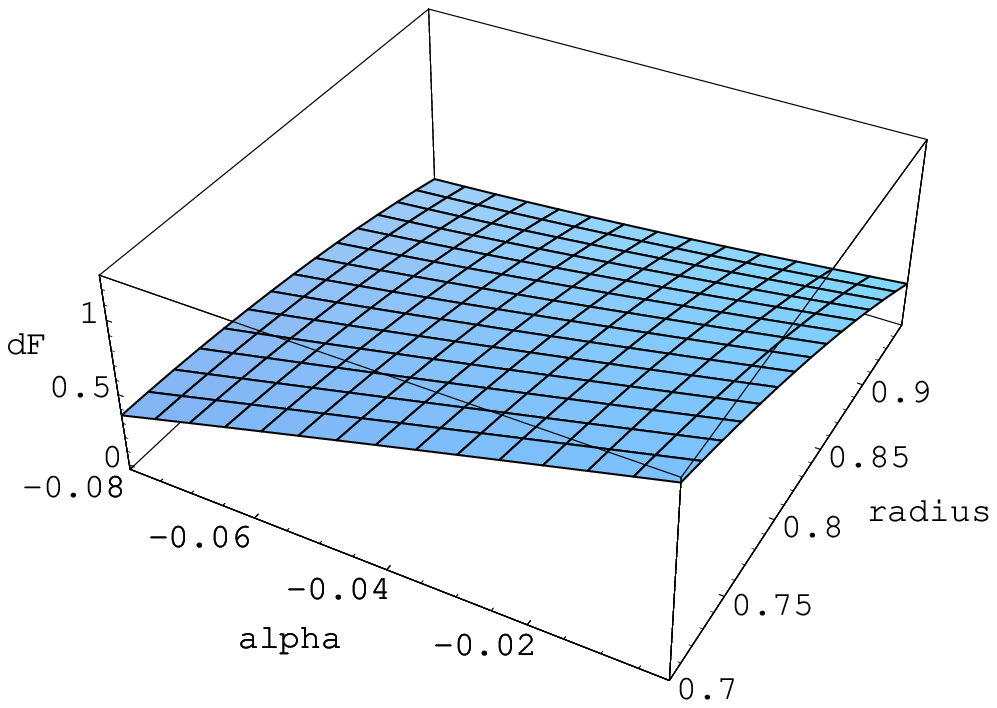,height=80mm,width=100mm,angle=0}
 \caption{The difference $\triangle F$ of two free energies in the
 case of five dimensions with $-1/12 <\tilde \alpha/l^2 <0$.}



\end{figure}



















\end{references}
\end{document}